\documentclass[aps,prc,twocolumn,showpacs,amsmath,amssymb,nofootinbib]{revtex4}
\usepackage {lscape}
\usepackage{graphicx}
\usepackage{amsmath}
\usepackage{rotating}

\usepackage[bookmarks=true]{hyperref}
\usepackage{color}
\usepackage{longtable}

  \hypersetup{
   colorlinks=true,
   pdfborder={0 0 0},
   citecolor=blue,
   linkcolor=blue,
   urlcolor=blue
 }

\begin{document}

\title{Spectroscopic factor strengths using $ab ~ initio$ approaches}

\author{P.C. Srivastava\footnote{pcsrifph@iitr.ac.in} and Vikas Kumar}

\address{Department of Physics, Indian Institute of Technology, Roorkee
247 667, India}

\date{\hfill \today}
\begin{abstract}

We have calculated the spectroscopic factor strengths for the one-proton and one-neutron pick-up reactions
$^{27}$Al($d$,$^{3}$He)$^{26}$Mg and 
$^{27}$Al($d$,$t$)$^{26}$Al within the framework of the shell model.  We employed two different
$ab ~ initio$ approaches : an in-medium similarity renormalization targeted for a particular nucleus, and the coupled-
cluster effective interaction. We also compared our results with recently determined experimental spectroscopic factors.

\end{abstract}
\maketitle
\section{Introduction}

The nature and occupancy  of the single-particle orbits  for a nucleus can be determined from the spectroscopic factors (SFs).
Experimentally the SFs can be measured with
single-particle transfer reactions. These reactions are of two types, first one is stripping in which one nucleon is
stripped from the incoming projectile, while the second one is pick-up reaction in which one nucleon is picked up by the projectile. 
The examples of neutron transfer pick-up reactions are $(p,d)$, $(d,t)$ and ($^{3}$He,$\alpha$),
while stripping reactions are $(d,p)$, $(t,d)$ and ($\alpha$, $^{3}$He)~\cite{bru}.
The SF is defined by a matrix element between initial and final state corresponding to entrance channel and exit channels,
respectively.
It is possible to describe the capture or emission of single nucleons in stellar burning processes by calculating the nuclear 
matrix elements
for  single-nucleon spectroscopic factors in the nuclear structure calculations. 

Studies of SFs in different region of nuclear chart are reported in Refs.
\cite{mh,sc1,sc2,brown,Tsang1,Franchoo,Lee}.
Survey of excited state neutron spectroscopic factors for $Z=8-28$ nuclei are reported by Tsang {\it et al} \cite{Tsang2}.
In this work they extracted 565 neutron spectroscopic factors for $sd$ and $fp$ shell nuclei by analyzing $(d,p)$ angular distributions,
they also compared the experimental results with shell-model results.

To study different excited states and their spectroscopic factors for $^{26}$Mg, many experimental
results were reported in Refs. \cite{mg1,mg2,mg3,mg4,mg5,mg6,mg7}. Recently, the spectroscopic factors for $^{26}$Mg is reported
in Ref. \cite{vishal1} using $^{27}$Al(d,$^{3}$He)$^{26}$Mg  reaction.
The structure of $^{27}$Al deduced from experiments which enlighten different reaction channels
were reported in refs. ~\cite{al1,al2,al3,al4,al5,al6,al7}. The experimental results for SFs
of 14 excited states for $^{27}$Al using $^{27}$Al(d,$t$)$^{26}$Al reaction are reported in Ref. \cite{vishal2}.
The studies of $^{26}$Al and $^{26}$Mg are important for astrophysics point of view.  The massive stars throughout the Galaxy
dominate in the production of $^{26}$Al ~\cite{nat}, and it decays by $\beta^+$ to $^{26}$Mg.

In the present work we have performed shell-model calculations using $ab ~ initio$ approaches for 
one-neutron and one-proton pick-up reaction on $^{27}$Al within the framework of the shell model.

\section{$Ab ~ initio$ Shell Model Analysis}

We performed shell-model calculations using two modern approaches: an in-medium similarity renormalization targeted for a particular nucleus \cite{rag2} and the
$ab ~ initio$ coupled-cluster effective interaction (CCEI) \cite{ccei2}. We also compared results with a phenomenological USDB 
interaction \cite{usdausdb}.
For the diagonalization of matrices we used shell-model code NuShellX \cite{nushellx}.

Recently,  Stroberg $\it {et ~ al.,}$ reported mass-dependent Hamiltonians for $sd$-shell nuclei using the
in-medium similarity renormalization group (IM-SRG) based on chiral two- and three-nucleon interactions \cite{rag1}.  Further extension has been done for
$ab~initio$ IM-SRG calculation 
based on ensemble reference states to consider residual $NNN$ forces among valence nucleons. 
This is a nucleus-dependent valence-space approach to study nuclear structure properties \cite{rag2}.
In the present work we performed calculations for the spectroscopic factor strengths using separate effective
interactions for $^{26}$Mg, $^{26}$Al and $^{27}$Al  based on nucleus-dependent valence space \cite{rag2}.

The $ab ~ initio$ coupled-cluster effective interaction (CCEI) \cite{ccei2}, uses A- dependent Hamiltonian,

\begin{equation}
  \label{intham}
  \hat{H} = \sum_{i<j}\left({({\bf p}_i-{\bf p}_j)^2\over 2mA} + \hat{V}
    _{NN}^{(i,j)}\right) + \sum_{ i<j<k}\hat{V}_{\rm 3N}^{(i,j,k)}.
\end{equation}

with an initial next-to-next-to-next-to-leading order ($N3LO$) chiral $NN$
interaction, and next-to-next-to-leading order ($N2LO$) local chiral $NNN$
interaction, using similarity 
renormalization group transformation. 
From this the shell-model Hamiltonian in a $sd$ valence-space obtained from $ab-initio$ coupled-cluster theory. 
Where, the CCEI Hamiltonian is 

\begin{equation}
  \label{ham}
  H_{\rm CCEI} = H_0^{A_c} + H_1^{A_c+1} + H_2^{A_c+2} + \cdots,
\end{equation}

For the shell-model space this Hamiltonian is limited for one- and two-body terms. The Two-body 
term is computed using Okubo-Lee-Suzuki similarity transformation. In the Eq. (2), A is the mass of
the nucleus, $A_c$ is the mass of core. $H_0^{A_c}$ is the Hamiltonian for the core, $H_1^{A_c+1}$ 
is the valence one-body Hamiltonian, and $H_2^{A_c+2}$ is the additional two-body Hamiltonian.
Coupled-cluster theory results for spectroscopic factor for proton and neutron removal from $^{16}$O is reported in Ref. \cite{nav}. 

The Hamiltonian of the USDB is based on a renormalized $G$ matrix by 
fitting two-body matrix elements with experimental data for binding energies and excitation energies 
for the $sd$ shell nuclei \cite{usdausdb,brown2}.
The USDB interactions is fitted by varying 56 linear combinations of two-body matrix elements.
The rms deviations of 130 keV were obtained between experimental and theoretical energies for USDB interaction.
In the present work before calculating the SF, first we examined the wave functions of concerned nuclei using
IM-SRG, CCEI and USDB effective interactions. 
The comparison between the calculated and experimental levels for $^{26}$Mg and $^{26}$Al is reported in Figs. 1 and 2.
The results of the USDB interaction are much better than IM-SRG and CCEI interactions. In the case of $^{26}$Al we 
calculated only the g.s.($0^+$).

We can define the spectroscopic amplitudes for pick-up and stripping reactions by taking
the expectation values of the operators $a^\dagger$ and $a$ between the states of nuclei with A-1 and A,
and A+1 and A. The spectroscopic factor in terms of the reduced matrix elements of $a^\dagger$ is given by: 

\begin{equation}
 {S} = \frac{1}{2J + 1} \, |{\langle \psi ^{A}\omega J ||a_{k}^\dagger||\psi ^{A-1}\omega{'} J^{'} \rangle}|^2,
\end{equation}

where the $(2J+1)$ factor is by convention associated with heavier mass $A$. Here, $\omega$ indices distinguish the various basis 
states with the same $J$ value \cite{poves,brownbook}.

Experimentally \cite{vishal1,vishal2}, the spectroscopic factors of different states were extracted using the relation between experimental
cross section and theoretical cross sections;

\begin{equation}
\mathrm{\left(\frac{d\sigma}{d\Omega}\right)_{exp.}}=3.33\frac{C^{2}S}{2J+1}\mathrm{\left(\frac{d\sigma}{d\Omega}\right)_{DWBA}},
\end{equation}

where, $\mathrm{(\frac{d\sigma}{d\Omega})_{exp.}}$ is the experimental
differential cross-section and $\mathrm{(\frac{d\sigma}{d\Omega})_{DWBA}}$ is
the cross-section predicted by the DWUCK4 code. $J$ ($J =\it l \pm \frac{1}{2}$) is the total angular momentum of
the orbital from where proton is picked up. $C^{2}$ is the isospin Clebsch-Gordon coefficient and  S is the spectroscopic factor.

The uncertainty in the experimental SFs may come due to following:
(i) the zero-range parameter $D_0$ may be uncertain, (ii) the optical potential may be uncertain, (iii) the zero-range distorted-wave Born-approximation
is not sufficient \cite{rad1,rad2,rad3}. In the Refs. \cite{vishal1,vishal2}, it was reported that the replacement of the zero-range approximation with finite-range and
nonlocal parameters reduces the SFs up to 45-50 \%. 

\begin{figure}[h]
\centering
\includegraphics[width=9.5cm,height=10cm,clip]{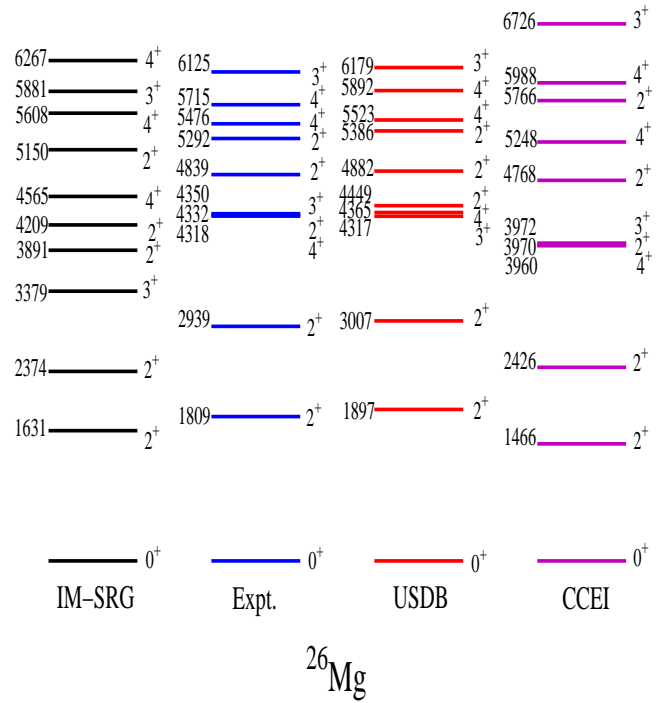}
\caption {\label{fig2}
Comparison between calculated energy levels and experimental data for $^{26}$Mg.} 
\end{figure}

\begin{figure}[h]
\centering
\includegraphics[width=9.5cm,height=10cm,clip]{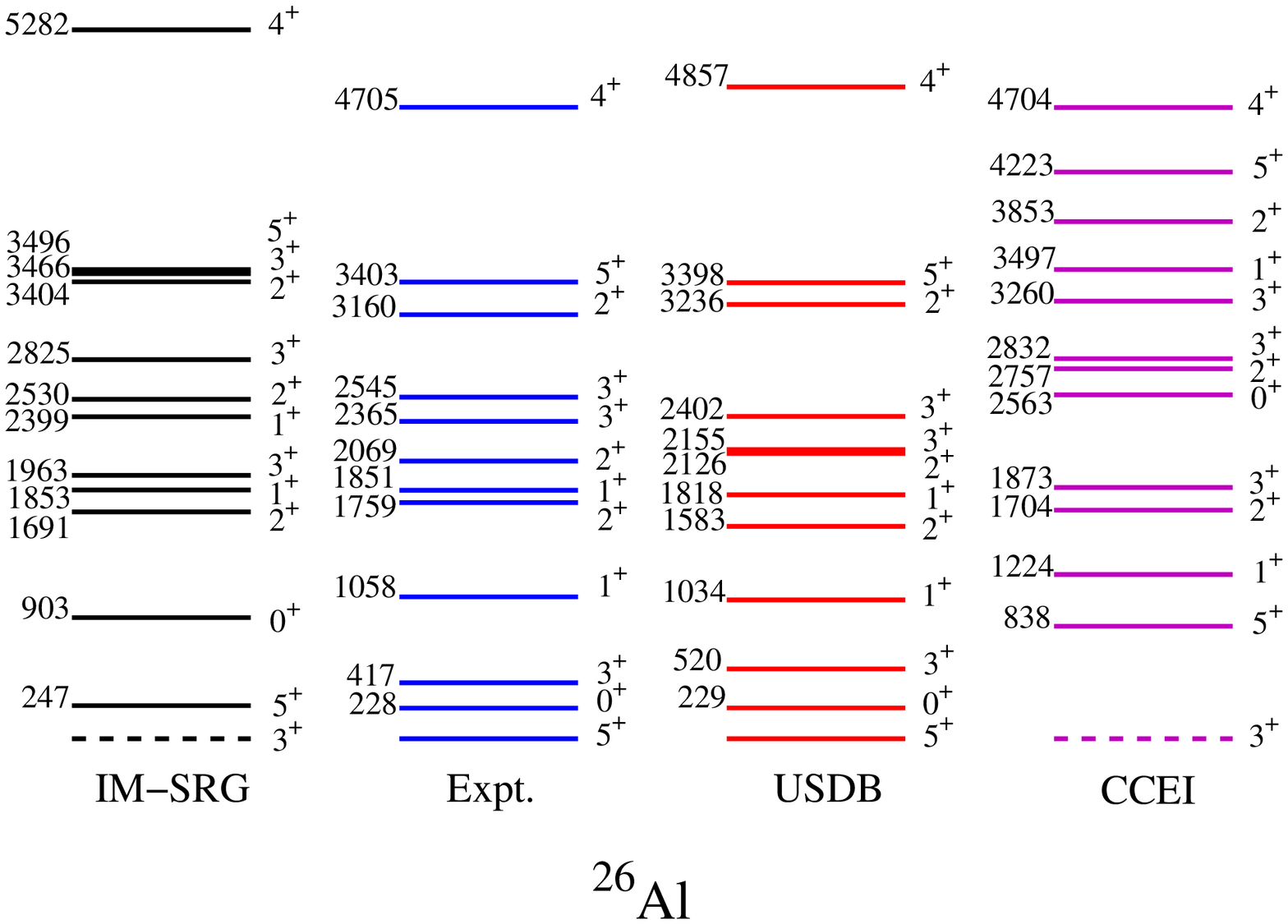}
\caption {\label{fig2}
 Comparison between calculated energy levels and experimental data for $^{26}$Al.} 
\end{figure}

Furnstahl and Hameer, using effective field theory, tried to determine whether occupation numbers and momentum
distributions of nucleons in nuclei are observables.  They claimed that these quantities
can only be defined if we take specific form of the Hamiltonian, regularization scheme etc. \cite{fur}. 
In the effective field theory, there is no definite form of the Hamiltonian, thus it is not possible to defined occupation numbers
(or even momentum distribution). The ``nonobservable" nuclear  quantities such as momentum distribution
and spectroscopic factor using parton distribution function ( PDFs) reported in Ref. \cite{fur_jpg}.
The inclusion of long-range (low-momentum) pion-exchange tensor forces is important.
But the recent study for the quenching of spectroscopic factors suggest that a long-range correlation is more dominating \cite{barbieri}. 
The uncertainty of the SFs coming from different sources was
reported in the review article by Dickhoff and Barbieri \cite{barbieri1}.  
In-elastic proton scattering is surface reaction, thus no detailed information is obtained related to the
interior of the nucleus; this will gives rise to an error of ~10\%.
Another uncertainty is due to the choice of the electron-proton cross-section; this will give a small uncertainty in the analysis of low- $Q^2$ data.

\begin{table*}
\centering
\caption{ \label{table1}  Extracted values of ${C^{2}S}$ for different excited states of $^{26}$Mg from the reaction $^{27}$Al(d,$^{3}$He) at 25 MeV.
We have taken experimental data from Ref. \cite{vishal1}. Here $l=0$ and $l=2$ are for $s_{1/2}$ and $d_{5/2}$ orbitals, respectively. }
   \begin{tabular}{rrrrcccccccccccccccccccccccc}
    \hline
     Expt.&{J$^{\pi}$} & [USDB]& [IM-SRG]& [CCEI] &&\multicolumn{2}{c}{${C^{2}S}$[Expt.]}
     && \multicolumn{2}{c}{${C^{2}S}$[USDB]}&& \multicolumn{2}{c}{${C^{2}S}$[IM-SRG]}&& \multicolumn{2}{c}{${C^{2}S}$[CCEI]} \\
     \cline{7-8}
     \cline{10-11}
     \cline{13-14}
     \cline{16-17}
keV& &keV& keV & keV &&\multicolumn{1}{c}{$l=0$}& \multicolumn{1}{c}{$l=2$}&&\multicolumn{1}{c}{$l=0$}& \multicolumn{1}{c}{$l=2$}&&
\multicolumn{1}{c}{$l=0$}& \multicolumn{1}{c}{$l=2$}&&\multicolumn{1}{c}{$l=0$}& \multicolumn{1}{c}{$l=2$}\\
\hline
0   &{0$^{+}_1$}& 0{(0$^{+}_1$)}     & 0{(0$^{+}_1$)}   &   0{(0$^{+}_1$)} &&       &0.17$\pm$0.05&&...  & 0.276 && ...   & 0.232&& ... & 0.339 \\ 
1806&{2$^{+}_1$}& 1897{(2$^{+}_1$)} & 1631{(2$^{+}_1$)}& 1466{(2$^{+}_1$)}&& 0.002 &0.57$\pm$0.14&&0.014& 0.876 && 0.006 &0.647 && 0.036 & 0.140 \\ 
2935&{2$^{+}_2$}& 3007{(2$^{+}_2$)} &2374{(2$^{+}_2$)} &2426{(2$^{+}_2$)} && 0.002 &0.13$\pm$0.03&&0.021& 0.090 && 0.030 & 0.072&& 0.002 &  0.170 \\ 
    &{${3}^{+}_1$}& 4317{(3$^{+}_2$)} &3379{(3$^{+}_1$)} &3972{(3$^{+}_1$)} &&       &              &&0.072& 0.005 && 0.0002&0.0001&& 0.055 & 0.000\\ 
    &{${4}^{+}_1$}& 4365{(4$^{+}_1$)} &4565{(4$^{+}_1$)} & 3960{(4$^{+}_1$)}&&    &              &&...  & 1.620 && ...   & 1.038&& ... &0.795\\ 
    &{${2}^{+}_3$}& 4449{(2$^{+}_3$)} & 3891{(2$^{+}_3$)}&3970{(2$^{+}_3$)} &&       &              &&0.044& 0.074 && 0.0002& 0.057&& 0.016 &0.196\\ 
    &{2$^{+}_4$}&  4882{(2$^{+}_4$)} & 4209{(2$^{+}_4$)}& 4768{(2$^{+}_4$)}&&       &               &&0.140& 0.022 && 0.220 &0.053 && 0.078 &0.091\\ 
    &{2$^{+}_5$}&  5386{(2$^{+}_5$)} & 5150{(2$^{+}_5$)}&5766{(2$^{+}_5$)} &&       &               &&0.009& 0.004 && 0.103 & 0.024&& 0.205 & 0.122\\ 
    &{4$^{+}_2$}&  5523{(4$^{+}_3$)} &5608{(4$^{+}_4$)} & 5248{(4$^{+}_3$)}&&       &               &&...  & 0.185 && ...   &0.030 && ... &0.009 \\ 
    &{4$^{+}_3$}&  5892{(4$^{+}_4$)} &6267{(4$^{+}_5$)} & 5988{(4$^{+}_4$)}&&       &               &&  ...& 0.001 && ...   & 0.030&& ... &0.025\\ 
    &{3$^{+}_2$}&  6179{(3$^{+}_3$)} &5881{(3$^{+}_3$)} &6726{(3$^{+}_3$)} &&       &               &&0.028& 0.006 && 0.067 & 0.022&& 0.185 &0.002\\      
\hline
\end{tabular}
\label{table2}
\end{table*}

\begin{table*}
\centering
\caption{ \label{table2} Wave functions of different states for $^{26}$Mg. The SFs corresponding to these
states are shown in the Table \ref{table1}.} 
    \begin{tabular}{rrrrccccccccccccccccccc}
    \hline

     &&\multicolumn{2}{c}{IM-SRG}&& \multicolumn{2}{c}{USDB}\\
     \cline{3-4}
     \cline{6-7}
 {J$^{\pi}$}  &&\multicolumn{1}{c}{$\%$}& \multicolumn{1}{c}{Configuration}&&\multicolumn{1}{c}{$\%$}& \multicolumn{1}{c}{Configuration}\\
\hline
{0$^{+}$}   && 49    &${\pi(d_{5/2})^4}\otimes{\nu(d_{5/2})^6}$                            &&  62  &${\pi(d_{5/2})^4}\otimes{\nu(d_{5/2})^6}$  \\ 
{2$^{+}$}   && 43    &${\pi(d_{5/2})^4}\otimes{\nu(d_{5/2})^6}$                            &&  52  &${\pi(d_{5/2})^4}\otimes{\nu(d_{5/2})^6}$   \\ 
{2$^{+}$}   && 29    &${\pi({d_{5/2}})^3({s_{1/2}})^1}\otimes{\nu({d_{5/2}})^5({s_{1/2}})^1}$&&  29  &${\pi(d_{5/2})^4}\otimes{\nu({d_{5/2}})^5({s_{1/2}})^1}$ \\ 
{3$^{+}$}   && 33    &${\pi(d_{5/2})^4}\otimes{\nu({d_{5/2}})^5({s_{1/2}})^1}$               &&  35  &${\pi(d_{5/2})^4}\otimes{\nu({d_{5/2}})^5({s_{1/2}})^1}$   \\ 
{4$^{+}$}   && 35    &${\pi(d_{5/2})^4}\otimes{\nu({d_{5/2}})^6}$                            &&  52  &${\pi(d_{5/2})^4}\otimes {\nu(d_{5/2})^6}$ \\ 
{2$^{+}$}   && 23    &${\pi(d_{5/2})^4}\otimes{\nu({d_{3/2}})^1({d_{5/2}})^4({s_{1/2}})^1}$  &&  25  &${\pi(d_{5/2})^4}\otimes {\nu(d_{5/2})^6}$ \\ 
{2$^{+}$}   && 28    &${\pi({d_{5/2}})^3({s_{1/2}})^1}\otimes{\nu({d_{5/2}})^6}$             &&  33  &${\pi({d_{5/2}})^3({s_{1/2}}})^1\otimes {\nu(d_{5/2})^6}$ \\ 
{2$^{+}$}   && 22    &${\pi(d_{5/2})^4}\otimes{\nu({d_{3/2}})^1({d_{5/2}})^5}$               &&  33  &${\pi(d_{5/2})^4}\otimes {\nu({d_{3/2}})^1({d_{5/2}})^5}$ \\ 
{4$^{+}$}   && 31    &${\pi(d_{5/2})^4}\otimes{\nu({d_{5/2}})^5({s_{1/2}})^1}$               &&  26  &${\pi(d_{5/2})^4}\otimes {\nu({d_{5/2}})^5({s_{1/2}})^1}$ \\ 
{4$^{+}$}   && 29    &${\pi(d_{5/2})^4}\otimes{\nu({d_{5/2}})^4({s_{1/2}})^2}$               &&  31  &${\pi(d_{5/2})^4}\otimes {\nu({d_{5/2}})^5({s_{1/2}})^1}$ \\ 
{3$^{+}$}   && 32    &${\pi(d_{5/2})^4}\otimes{\nu({d_{3/2}})^1({d_{5/2}})^5}$               &&  50  &${\pi(d_{5/2})^4}\otimes {\nu({d_{5/2}})^5({s_{1/2}})^1}$ \\
\hline

\end{tabular}
\label{table2}
\end{table*}

\begin{table*}
\centering
\caption{ \label{table3} Extracted values of ${C^{2}S}$ for different excited states of $^{26}$Al from the reaction $^{27}$Al$(d,t)$.
We have taken experimental data from Ref. \cite{vishal2}. Here $l=0$ and $l=2$ are for $s_{1/2}$ and $d_{5/2}$ orbitals, respectively.} 
     \begin{tabular}{rrrrcccccccccccccccccccccccc}
    \hline
     Expt.&{J$^{\pi}$} & [USDB]& [IM-SRG]& [CCEI] &&\multicolumn{2}{c}{${C^{2}S}$[Expt.]}
     && \multicolumn{2}{c}{${C^{2}S}$[USDB]}&& \multicolumn{2}{c}{${C^{2}S}$[IM-SRG]}&& \multicolumn{2}{c}{${C^{2}S}$[CCEI]} \\
     \cline{7-8}
     \cline{10-11}
     \cline{13-14}
     \cline{16-17}
keV& &keV& keV & keV &&\multicolumn{1}{c}{$l=0$}& \multicolumn{1}{c}{$l=2$}&&\multicolumn{1}{c}{$l=0$}& \multicolumn{1}{c}{$l=2$}&&
\multicolumn{1}{c}{$l=0$}& \multicolumn{1}{c}{$l=2$}&&\multicolumn{1}{c}{$l=0$}& \multicolumn{1}{c}{$l=2$}\\
\hline
0     &{5$^{+}_1$}&0{(5$^{+}_1$)}    &247{(5$^{+}_1$)} &838{(5$^{+}_1$)} && ... &  0.73$\pm$0.21   &&   ... & 1.089 && ...   &0.811 &&...   & 0.627 \\ 
228.3 &{0$^{+}_1$}& 229{(0$^{+}_1$)} &903{(0$^{+}_1$)} &2563{(0$^{+}_2$)} && ... &0.09$\pm$0.03    &&   ... & 0.139 &&...    &0.117 && ...  & 0.004\\ 
416.8 &{3$^{+}_1$}& 520{(3$^{+}_1$)} &1963{(3$^{+}_2$)}&1873{(3$^{+}_3$)} && ... &  0.32$\pm$0.07  && 0.201 & 0.0001&& 0.0003& 0.321&&0.028 & 0.004\\ 
1057.7&{1$^{+}_1$}& 1034{(1$^{+}_1$)}&1853{(1$^{+}_3$)}&1224{(1$^{+}_3$)} && ... &   0.17$\pm$0.05 && ...  &0.004  &&...    &0.0007&&...   & 0.114\\ 
1759.0&{2$^{+}_1$}& 1583{(2$^{+}_1$)}& 1691{(2$^{+}_2$)}&1704{(2$^{+}_3$)} && ... &  0.038$\pm$0.006&& 0.031  &0.004  && 0.001 &0.0001&&0.046 & 0.023\\ 
1850.6&{1$^{+}_2$}& 1818{(1$^{+}_2$)}&2399{(1$^{+}_4$)}&3497{(1$^{+}_5$)} && ... &   0.019$\pm$0.004&& 0.007  & 0.438 &&...    & 0.001&& ...  & 0.027 \\ 
2068.8&{2$^{+}_2$}& 2126{(2$^{+}_2$)}&2530{(2$^{+}_3$)}&2757{(2$^{+}_4$)} && ... &    0.26$\pm$0.06&& 0.007  & 0.438 && 0.0004&0.002 &&0.007 & 0.146\\ 
2365.1&{3$^{+}_2$}& 2155{(3$^{+}_2$)}&2825{(3$^{+}_4$)}& 2832{(3$^{+}_5$)}&& ... &   0.13$\pm$0.02  && 0.008  & 0.310 &&0.0005 & 0.005&& 0.022& 0.071\\ 
2545.3&{3$^{+}_3$}& 2402{(3$^{+}_3$)}&3466{(3$^{+}_5$)}&3260{(3$^{+}_6$)} && ... & 0.16$\pm$0.03    && 0.002  & 0.120 && 0.001 & 0.008&& 0.049& 0.068 \\ 
3159.8&{2$^{+}_3$}& 3236{(2$^{+}_5$)}&3404{(2$^{+}_6$)}&3853{(2$^{+}_6$)} && ... &  0.06$\pm$0.01   && 0.010  & 0.045 &&0.016  & 0.051&& 0.000&0.068 \\ 
3402.6&{5$^{+}_2$}& 3398{(5$^{+}_2$)}& 3496{(5$^{+}_2$)}&4223{(5$^{+}_3$)} && ... &0.06$\pm$0.01    &&   ...  & 0.077 &&...    & 0.133&& ...  & 0.0007\\ 
4705.3&{4$^{+}_1$}& 4857{(4$^{+}_5$)}& 5282{(4$^{+}_5$)}&4704{(4$^{+}_5$)} && ... &0.27$\pm$0.08    &&   ...  & 0.0003&& ...   &0.007 && ...  & 0.008\\     
\hline
\end{tabular}
\label{table2}
\end{table*}

\begin{figure*}
\centering
\includegraphics[width=13cm,height=10cm,clip]{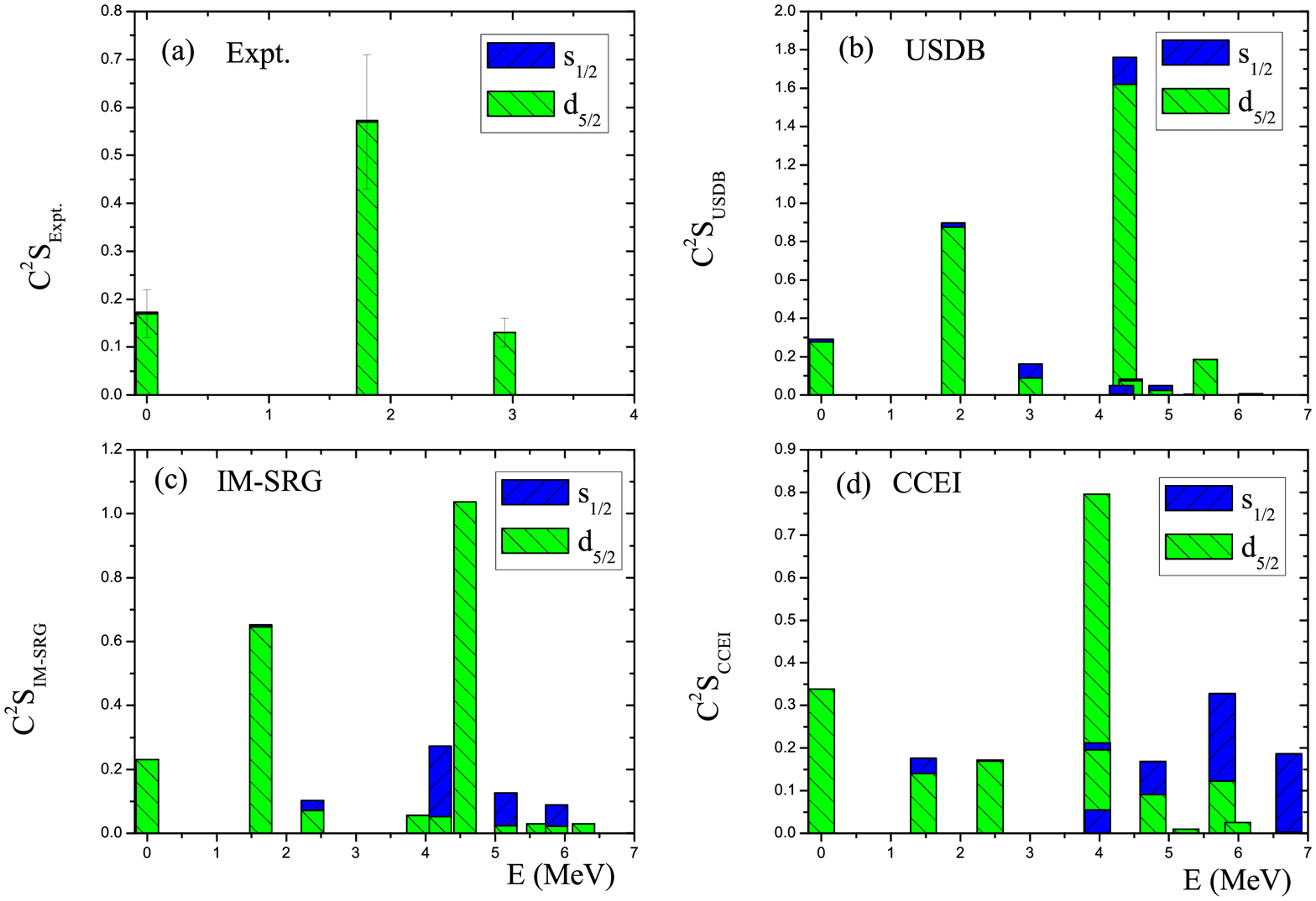}
\includegraphics[width=10cm,height=7cm,clip]{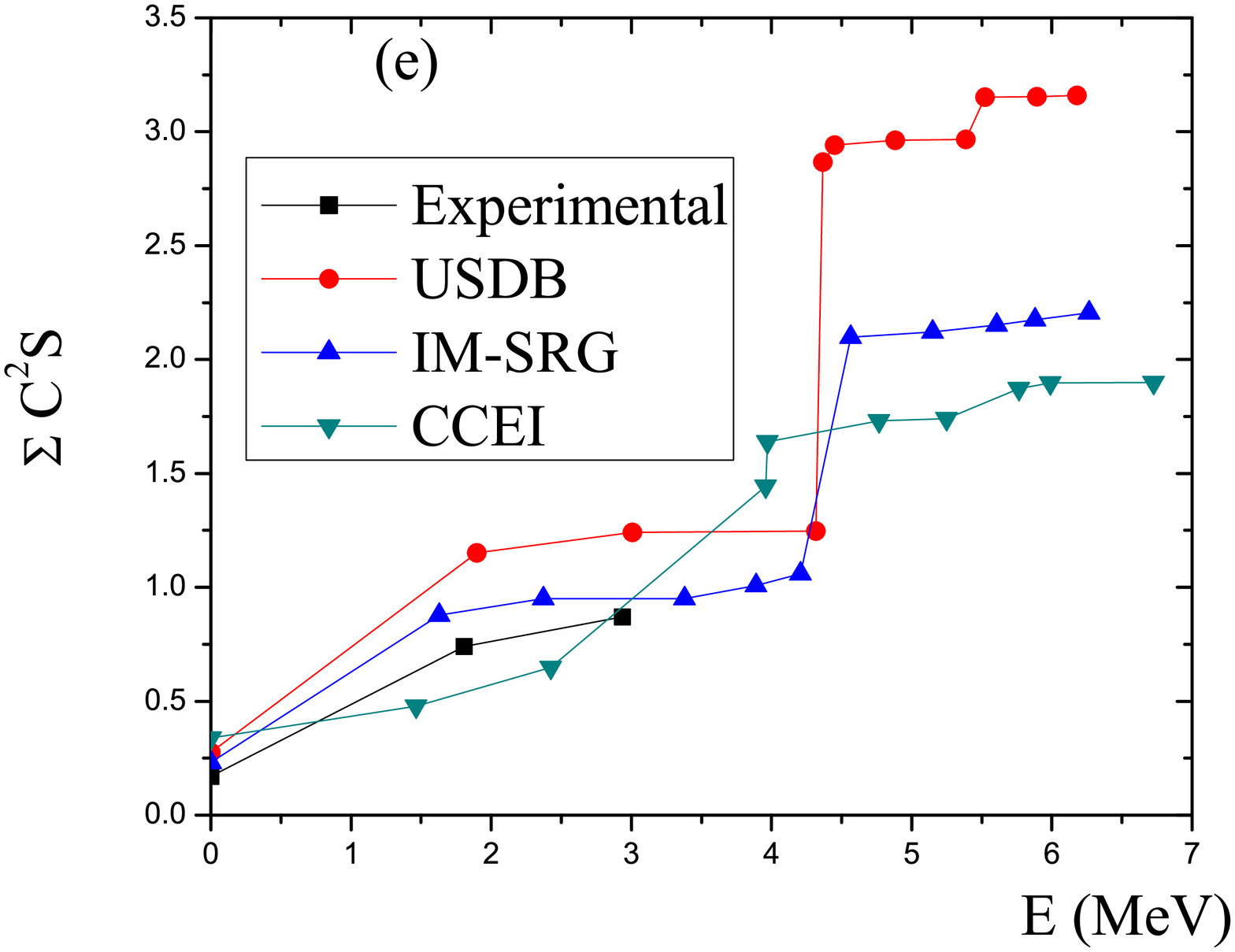}
\caption { $C^2S$ comparison between calculated and extracted values from experimental data using the zero-range DWBA-model calculation
 \cite{vishal1} for $^{26}$Mg.} 
\end{figure*}

\begin{figure*}
\centering
\includegraphics[width=13cm,height=10cm,clip]{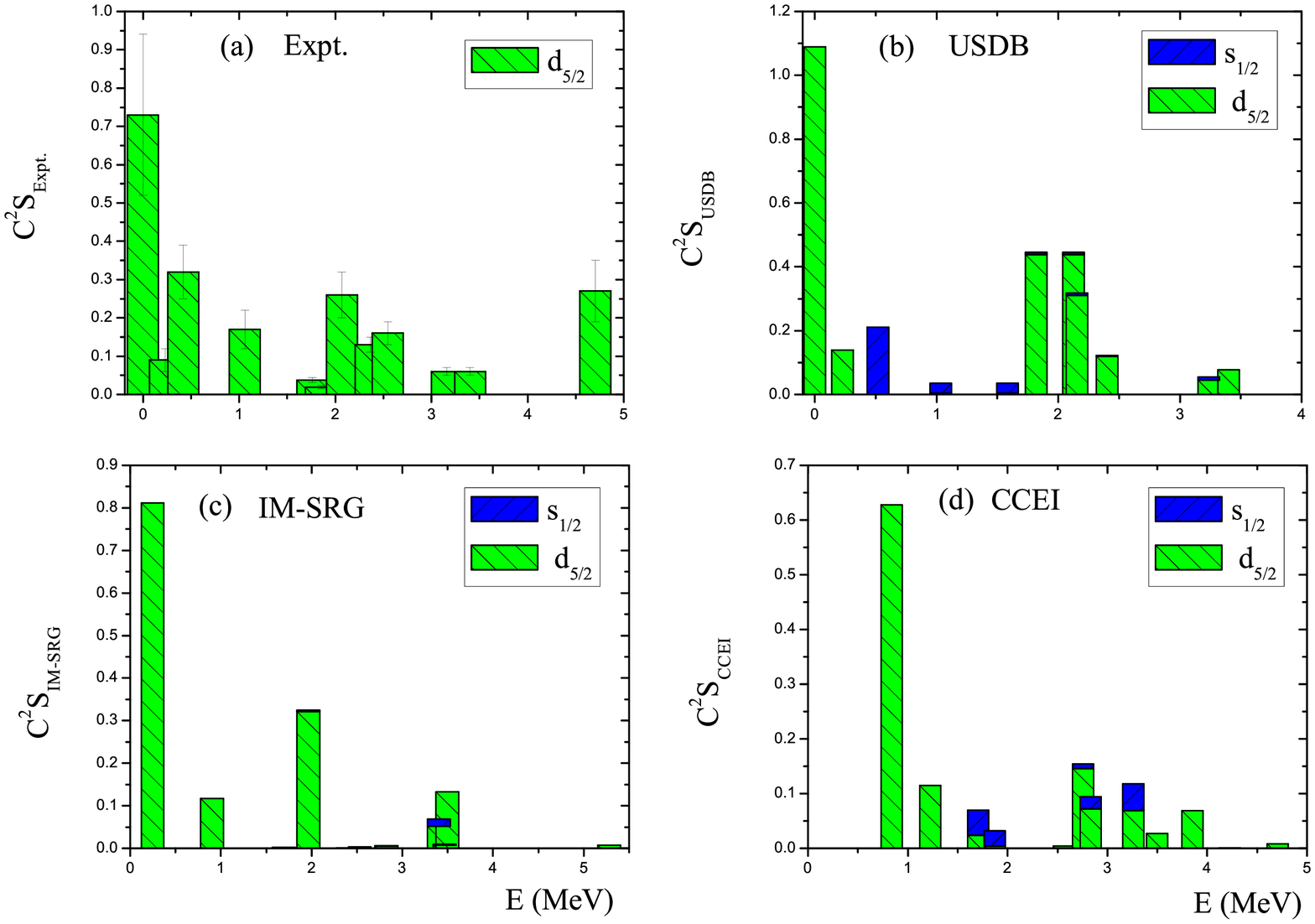}
\includegraphics[width=10cm,height=7cm,clip]{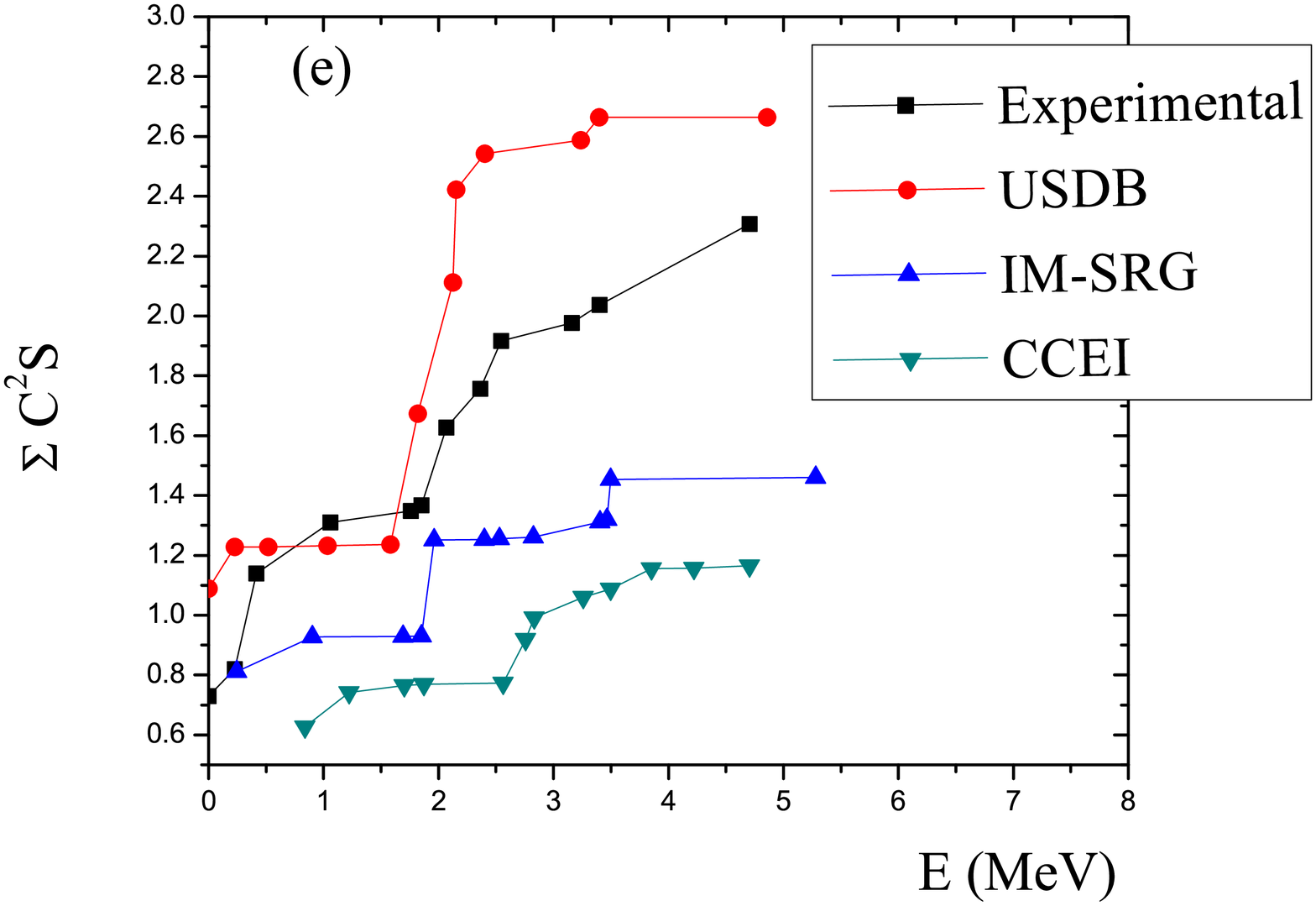}
\caption {  $C^2S$ comparison between calculated and extracted values from experimental data using the zero-range DWBA-model calculations
\cite{vishal2} for $^{26}$Al. } 
\end{figure*}

\begin{table*}
\centering
\caption{ \label{table4}  Wave functions of different states for $^{26}$Al. The spectroscopic factors corresponding to these
states are shown in the Table \ref{table3}.}
    \begin{tabular}{rrrrccccccccccccccccccc}
    \hline

    &&\multicolumn{2}{c}{IM-SRG}&& \multicolumn{2}{c}{USDB}\\
     \cline{3-4}
     \cline{6-7}
 {J$^{\pi}$} &&\multicolumn{1}{c}{$\%$}& \multicolumn{1}{c}{Configuration}&&\multicolumn{1}{c}{$\%$}& \multicolumn{1}{c}{Configuration}\\
\hline
{5$^{+}$}   && 48    &${\pi(d_{5/2})^5}\otimes{\nu(d_{5/2})^5}$                           &&  62  &${\pi(d_{5/2})^5}\otimes{\nu(d_{5/2})^5}$   \\ 
{0$^{+}$}   && 54    &${\pi(d_{5/2})^5}\otimes{\nu(d_{5/2})^5}$                           &&  65  &${\pi(d_{5/2})^5}\otimes{\nu(d_{5/2})^5}$    \\ 
{3$^{+}$}   && 23    &${\pi({d_{5/2}})^4({s_{1/2}})^1}\otimes{\nu({d_{5/2}})^4({s_{1/2}})^1}$&&  18  &${\pi({d_{5/2}})^4({s_{1/2}})^1}\otimes{\nu(d_{5/2})^5}$  \\ 
{1$^{+}$}   && 15    &${\pi({d_{5/2}})^4({s_{1/2}})^1}\otimes{\nu({d_{5/2}})^4({s_{1/2}})^1}$&&  57  &${\pi(d_{5/2})^5}\otimes{\nu(d_{5/2})^5}$   \\ 
{2$^{+}$}   && 13    &${\pi({d_{3/2}})^1({d_{5/2}})^4}\otimes{\nu({d_{5/2}})^4({s_{1/2}})^1}$&&  16  &${\pi({d_{3/2}})^1({d_{5/2}})^4}\otimes{\nu({d_{5/2}})^5}$ \\ 
{1$^{+}$}   && 16    &${\pi({d_{5/2}})^4({s_{1/2}})^1}\otimes{\nu({d_{5/2}})^4({s_{1/2}})^1}$&&  19  &${\pi({d_{5/2}})^4({s_{1/2}})^1}\otimes{\nu({d_{5/2}})^4({s_{1/2}})^1}$\\ 
{2$^{+}$}   && 14    &${\pi({d_{3/2}})^1({d_{5/2}})^4}\otimes{\nu(d_{5/2})^5}$               &&  42  &${\pi(d_{5/2})^5}\otimes{\nu(d_{5/2})^5}$  \\ 
{3$^{+}$}   && 17    &${\pi({d_{3/2}})^1({d_{5/2}})^4}\otimes{\nu({d_{3/2}})^1({d_{5/2}})^4}$&&  30  &${\pi(d_{5/2})^5}\otimes{\nu(d_{5/2})^5}$  \\ 
{3$^{+}$}   && 14    &${\pi({d_{3/2}})^1({d_{5/2}})^4}\otimes{\nu(d_{5/2})^5}$               &&  21  &${\pi(d_{5/2})^5}\otimes{\nu(d_{5/2})^5}$  \\ 
{2$^{+}$}   && 18    &${\pi({d_{5/2}})^5}\otimes{\nu({d_{5/2}})^4({s_{1/2}})^1}$             &&  13  &${\pi({d_{5/2}})^4{(s_{1/2}})^1}\otimes{\nu({d_{5/2}})^4({s_{1/2}})^1}$\\ 
{5$^{+}$}   && 16    &${\pi(d_{5/2})^5}\otimes{\nu(d_{5/2})^5}$                              &&  17  &${\pi(d_{5/2})^5}\otimes{\nu(d_{5/2})^5}$  \\
{4$^{+}$}   && 24    &${\pi({d_{5/2}})^5}\otimes{\nu({d_{5/2}})^4({s_{1/2}})^1}$             &&  17  &${\pi(d_{5/2})^5}\otimes{\nu({d_{5/2}})^4({s_{1/2}})^1}$  \\
\hline

\end{tabular}
\label{table2}
\end{table*}

\subsection{Calculation of ${C^{2}S}$ for $1p$ pick-up reaction $^{27}$Al(d,$^{3}$He)$^{26}$Mg}
 Experimentally the states of $^{26}$Mg \cite{vishal1} were studied
by assuming pick-up from $d_{5/2}$ orbital only, and also a few states by assuming configuration mixing 
of two lower orbital of $sd$ shell: $d_{5/2}$ and $s_{1/2}$ single particle orbitals.
In Table \ref{table1}, we compared the experimental ${C^{2}S}$ values with shell-model results for IM-SRG, CCEI and USDB interactions
(the corresponding wave functions are shown in Table {\color {blue} II} ).
As extracted from experiment the calculated  ${C^{2}S}$ values were very large for 
$l=2$ ($d_{5/2}$) transfer as compared with $l=0$ ($s_{1/2}$) transfer. The shell-model results are larger for first two states, 
thus assigning larger single particle characteristics to these 
states. In the present work we have also predicted ${C^{2}S}$ values for states up to
$\sim$ 6 MeV.
In the Fig. 3(a)-(d) we show the variation of ${C^{2}S}$ for extracted experiment and calculated values. In all the three shell-model calculations,
the spectroscopic factor for pickup from the $s_{1/2}$ orbital plays a major role for the higher excited states.
In the Fig. 3(e), we have also plotted $\sum{C^{2}S}$ values
for theory and extracted experimental value. The $\sum{C^{2}S}$ values calculated from IM-SRG and CCEI interactions are showing same trends
as extracted values from the experiment.

\subsection{Calculation of ${C^{2}S}$ for $1n$ pick-up reaction $^{27}$Al(d,$t$)$^{26}$Al}

The states of $^{26}$Al \cite{vishal2} were experimentally  studied
by assuming pick-up from $d_{5/2}$ and $s_{1/2}$ single particle orbitals,
 while $6^+$ state at 3507 keV by pick-up from $g_{9/2}$ orbital.
In the Table \ref{table3}, we compared experimental ${C^{2}S}$ values with shell-model results for IM-SRG, CCEI and USDB
interactions. The experimentally extracted value for spectroscopic factors up to $\sim$ 4.7 MeV are
reported in Ref. \cite{vishal2}. In the present work we interpreted these extracted SFs in term of shell-model calculations
(the corresponding wave functions are shown in Table \ref{table4}).
The experimental ${C^{2}S}$ values for $d_{5/2}$ states are given in the Table \ref{table3}.  For the  $5^+$ state the SF result with IM-SRG and USDB is 
slightly higher than extracted experimental value, while it is smaller with CCEI. 
For some states the SFs are very small this is because 
in these cases the wave functions are very fragmented.
This is because large cancellations of contributions from different components of the wave functions.
For $2^+$ at 1759 keV and $4^+$ at 4705 keV, the IM-SRG, CCEI and USDB interactions are predicting very small value of spectroscopic factors. 

In the Figs. 4(a)-(d) we show the variation of ${C^{2}S}$ for extracted experiment and calculated values. In the Fig. 4(e), we have also plotted $\sum{C^{2}S}$ values
for theory and extracted experimental value.
The extracted experimental  $\sum{C^{2}S}$ values show a good trend with IM-SRG and CCEI.

\section{Summary}

We performed shell-model calculations for spectroscopic factors with two $ab~initio$ approaches:
an in-medium similarity renormalization targeted for a particular nucleus and
coupled-cluster effective interaction (CCEI). We also performed calculations with realistic USDB effective interaction.
Along with the  $ab~initio$ results, we present a comparison with recently determined experimental spectroscopic factors.



\section*{ACKNOWLEDGMENTS}
              


P.C.S. thanks P. Navr\'atil and S. R. Stroberg for useful
discussions during this work. We also thank R. Shyam
and V. K. B. Kota for useful comments to improve this
manuscript. P.C.S. acknowledges the hospitality extended to
him during his stay at TRIUMF, Vancouver City, Canada.
P.C.S. acknowledges financial support from faculty initiation
grants. V.K.’s work was supported in part by the CSIR
GrantNo.09/143(0844)/2013-EMR-1-India Ph.D.fellowship
program

\end{document}